\documentclass[%
 aip,
 amsmath,amssymb,
 reprint,%
]{revtex4-1}

\usepackage{graphicx}
\usepackage{dcolumn}
\usepackage{bm}

\usepackage[utf8]{inputenc}
\usepackage[T1]{fontenc}
\usepackage{mathptmx}
\usepackage{etoolbox}

\makeatletter
\def\@email#1#2{%
 \endgroup
 \patchcmd{\titleblock@produce}
  {\frontmatter@RRAPformat}
  {\frontmatter@RRAPformat{\produce@RRAP{*#1\href{mailto:#2}{#2}}}\frontmatter@RRAPformat}
  {}{}
}%
\makeatother
\begin{document}

\preprint{AIP/123-QED}

\title{Machine-Learning Optimization and Characterization of a High–Optical-Depth Two-Color Nanofiber Trap}

\author{W. Crump}
\author{M. Sadeghi}%
\author{M.D. Hoogerland}
 \homepage{https://profiles.auckland.ac.nz/m-hoogerland}
 \email{m.hoogerland@auckland.ac.nz}
\affiliation{{Department of Physics and Dodd-Walls Centre for Photonic and Quantum Technologies,
 University of Auckland,  Private Bag 92019,  Auckland,  New Zealand}}




\begin{abstract}
Optical nanofibers provide a way of coupling quantum information in cold atoms across large distances, however, this coupling requires atoms to reside close to the nanofiber surface. Atoms can be trapped close to the surface using a two-color dipole trap. Here we present our experimental realization of a two-color dipole trap. We optimize the number of trapped atoms using a machine learning algorithm and measure the optical density via the transmission. We estimate the number of atoms in the trap to be approximately 1400 and the lifetime of the atoms in the trap to be 28 ms. Machine-learning optimization improved the on-resonance optical depth from 0.5 in the initial optimization stage to optical depths exceeding 15.
\end{abstract}

\flushbottom
\maketitle

\thispagestyle{empty}

\section*{Introduction}

The realization of scalable quantum networks requires efficient and coherent interfaces between light and matter \cite{Kimble08}. Laser-cooled atoms are particularly attractive quantum nodes: they offer long coherence times, well-controlled internal level structures, and strong optical transitions suitable for quantum state preparation, storage, and readout. To distribute quantum information, these atomic systems must be coupled efficiently to guided photonic modes that allow photons to act as carriers of quantum information over long distances.

Optical nanofibers provide a powerful architecture for achieving this interface. When a standard optical fiber is adiabatically tapered to subwavelength diameter, a significant fraction of the guided optical mode propagates outside the silica surface as an evanescent field. Cold atoms positioned within a few hundred nanometers of the fiber can couple strongly to these guided modes. Early work demonstrated the efficient coupling of atomic fluorescence into nanofiber modes \cite{nayak2007optical,sague2007cold}, establishing nanofibers as versatile tools for manipulating and probing cold atoms. Since then, nanofiber-based platforms have enabled collective strong coupling in fiber cavities \cite{Ruddell2017CollectiveStrongCoupling}, strong coupling to single trapped atoms \cite{Kato2015StrongCouplingAllFiber}, real-time observation of single atoms in nanofiber cavities \cite{Nayak2019RealTimeNanofiber}, collective excitations of atomic arrays coupled to waveguides \cite{Corzo2019Waveguide}, generation of Rydberg atoms near fiber surfaces \cite{PhysRevResearch.2.012038}, and photon-mediated interactions between distant atoms in coupled-cavity quantum electrodynamics systems \cite{Kato2019DressedStates}.

The coupling strength between an atom and the guided mode depends critically on the local intensity of the evanescent field \cite{le_kien2005_spontaneous}. Stronger coupling requires atoms to reside close to the fiber surface, but atoms in this regime experience attractive van der Waals forces that pull them toward the dielectric surface, leading to atom loss and decoherence. A practical solution is the implementation of a two-color evanescent-field dipole trap \cite{le_kien2004_atom_trap_waveguide,le_kien2005_state_insensitive_trap}. In this scheme, a red-detuned field provides an attractive potential, while a blue-detuned field generates a repulsive barrier near the surface. Because the evanescent decay length depends on wavelength, the combination of these two fields creates a trapping minimum at a well-defined distance from the fiber surface. Using counter-propagating red-detuned beams forms a standing wave along the fiber axis, generating a one-dimensional lattice of trapping sites.

The first experimental realization of such a nanofiber-based two-color trap was reported in 2010 \cite{vetsch2010_optical_interface}. Subsequent work demonstrated state-insensitive (“magic”) trapping configurations that cancel scalar and vector light shifts \cite{lacroute2012_state_insensitive_nanofiber_trap,goban2012_state_insensitive_nanofiber}, significantly improving the coherence properties of the atoms. More recently, nanofiber traps with sub-wavelength lattice spacing have been demonstrated \cite{pache2025_magic_wavelength_nanofiber}, opening new possibilities for engineered atomic arrays with tailored collective properties. Polarization control of the nanofiber modes has also been refined using imaging techniques \cite{Tkachenko2019PolarisationControl}, which are crucial for optimizing trapping geometries. Comprehensive theoretical descriptions of nanofiber mode structures and atom–field interactions are provided in Refs. \cite{Nayak2018NanofiberQuantumPhotonics,Vetsch2011NanofiberInterface}.

Despite these advances, the implementation of a robust, high optical-depth nanofiber trap remains experimentally demanding. Trap depth, polarization configuration, laser noise, and cooling sequence parameters are strongly interdependent, and efficient loading into shallow near-surface traps is highly sensitive to experimental timing. Recently, machine-learning–based optimization techniques have been applied to nanofiber dipole traps \cite{gupta2022_machine_learner_opt}, building on earlier demonstrations of machine-learning control in ultracold-atom experiments \cite{Wigley2016FastMLOO}. The M-LOOP optimization framework \cite{mloop} provides a flexible platform for such automated tuning. However, the extent to which such approaches provide insight into the underlying parameter landscape, rather than simply improving performance, remains largely unexplored.

Here we report on the implementation and characterization of a two-color dipole trap for cesium atoms around an optical nanofiber in our laboratory. Using a blue-detuned field at 685 nm and counter-propagating red-detuned fields at 937 nm \cite{le_kien2005_state_insensitive_trap}, we generate an axial lattice of trapping sites located approximately 350 nm from the fiber surface with a calculated trap depth of $\sim 100~\mu$K. To efficiently load atoms into this relatively shallow near-surface trap, we integrate machine-learning–based optimization of a 17-parameter cooling sequence. Following optimization, we measure a peak optical depth of 15.1 $\pm$ 0.3 using transmission spectroscopy, corresponding to approximately 1400 trapped atoms and a collective cooperativity of $C_N \approx 28$. We further characterize trap frequencies, lifetime, and probe-induced heating dynamics.

While machine-learning optimization has previously been applied to nanofiber-based dipole traps \cite{gupta2022_machine_learner_opt}, its role has largely been limited to improving loading efficiency without detailed analysis of the underlying parameter landscape or experimental limitations. In contrast, we use machine learning both as an optimization tool and to identify experimentally sensitive parameter combinations, particularly the interplay between magnetic field control and cloud positioning
This enables efficient loading into a shallow trap configuration where manual optimization is challenging due to the strong coupling between experimental parameters.

In addition, by combining measurements of trap frequencies, optical depth, and probe-induced heating, we identify technical noise and anisotropic confinement as the primary mechanisms limiting the trap lifetime, rather than fundamental processes such as recoil heating or background gas collisions. This provides a quantitative understanding of the current performance limitations and defines clear pathways for improvement. 

These results establish a reproducible, high–optical-depth nanofiber interface in a regime of relatively weak confinement, and demonstrate that machine-learning–based optimization can be used to reach and diagnose this regime efficiently. The system is therefore well suited for future studies of collective emission, photon-mediated feedback, and waveguide quantum electrodynamics.

\section*{Experimental Setup}

\begin{figure}
\centering
	\includegraphics[width=\linewidth]{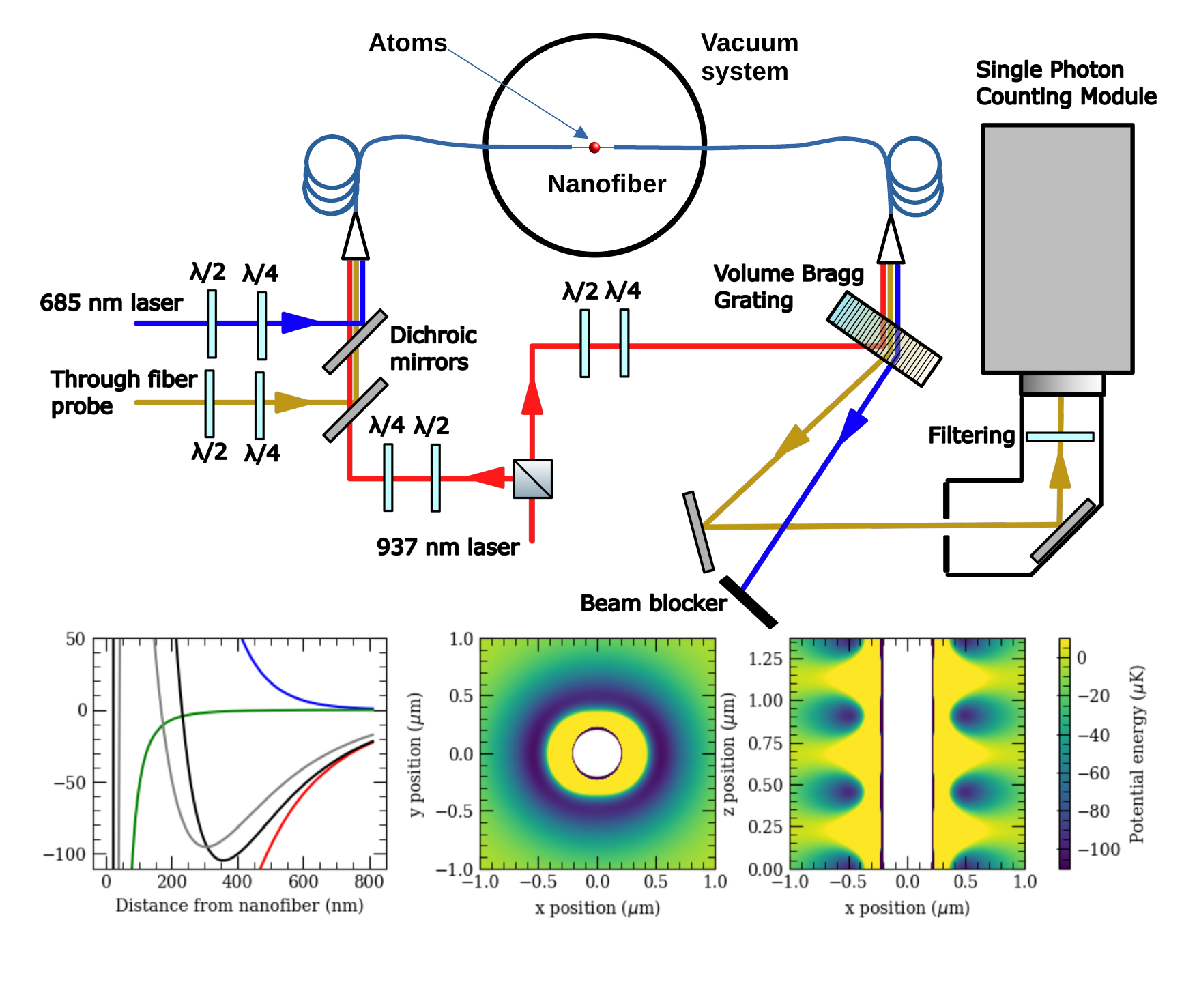} 
	\caption{ {\bf (Top)} Configuration of dipole trap lasers through the optical nanofiber as well as measurement probe with the SPCM. {\bf(Bottom Left)} Contributions to the potential energy from 685 nm laser (blue), 937 nm lasers (red), Van der Waals potential (green) and their sum (black) for a slice at $y=0$ and $z=0$ along the $x$ direction showing a trap depth of roughly 105 uK at a distance of around 350 nm from the fiber surface. A slice along the $y$ direction is also shown (grey). {\bf(Bottom Middle)} Potential energy landscape in the $x$-$y$ plane calculated using powers and polarizations from the experiment. {\bf(Bottom Right)} Potential energy landscape in the $x$-$z$ plane calculated using powers and polarizations from the experiment.}
	\label{fig:experimentalsetup}
\end{figure}

Our experimental setup is similar to that used in Refs. \cite{Ruddell2017CollectiveStrongCoupling,Sadeghi26}. Briefly, we create an optical nanofiber by simultaneously heating and stretching a standard single-mode fiber \cite{Ward2014NanofiberRig}. The 3~mm-long waist has a diameter of $\sim 380$~nm with an adiabatic transition between the core of the standard fiber to the nanofiber. The nanofiber is suspended at the center of a vacuum chamber. The top of figure \ref{fig:experimentalsetup} shows the configuration of the lasers going through the optical fiber. A 685~nm laser, blue-detuned from the Cs 
(6s)$\rightarrow$(6p) $D_2$
transition (which is at 852~nm) is used to create the repulsive potential for our two-color trap, with a power of 9.2 mW. Two counter-propagating, red-detuned 937 nm lasers are used for the attractive potential of the two-color trap. Each of these beams has a power of 260 $\mu$W through the nanofiber and together they create an array of trapping sites due to their standing wave configuration. These so-called "magic" wavelengths were used as they cancel out scalar light shifts caused by the off-resonant trapping lasers \cite{le_kien2005_state_insensitive_trap}.

The dipole trapping lasers both have the same linear polarization at the nanofiber, which is controlled using a half-- and quarter wave-plate for each beam. These polarizations are analyzed using a camera with a linear polarizer aligned to allow perpendicular polarizations to the fiber axis \cite{goban2012_state_insensitive_nanofiber,Tkachenko2019PolarisationControl}. The camera images the Rayleigh scattered light from the nanofiber and the half-- and quarter wave plates are rotated to find the maxima or minima in the detected light. These correspond to the linear polarized states in the direction of the linear polarizer or in the axis the camera points along. We use a Volume Bragg Grating (VBG) to separate the different wavelengths on one side of the nanofiber, which efficiently suppresses the trapping frequencies going into the SPCM.  

The lower plots of figure \ref{fig:experimentalsetup} show the calculated potential energy landscape due to the two-color trap \cite{le_kien2004_atom_trap_waveguide,Nayak2018NanofiberQuantumPhotonics,Vetsch2011NanofiberInterface}. An array of trapping sites are created along the fiber axis due to the 937 nm standing wave, which determines that the spacing of the trapping sites is 468.5~nm. 

Because of the polarizations we use, the atoms are not so well confined along the azimuthal direction indicated in figure \ref{fig:experimentalsetup}. The bottom left graph shows the depth along cuts in the $x$ and $y$ directions, which differ in depth of around 10 $\mu$K. This can also be seen in the trap frequencies calculated using $\omega_i = |\sqrt{\frac{1}{m}\frac{\partial^2U}{\partial x_I^2}}|$. For the radial and $z$ directions these are 609 kHz and 1.02 MHz (which gives a Lamb-Dicke parameter of $\sim$0.1 for our probe laser).

Much better confinement can be found in using orthogonal linear polarizations for the 685 nm and 937 nm lasers, but we were limited here by the maximum power of our 685 nm laser. If we had used this perpendicular configuration, our trap depth would be much smaller (at least half) due to having to ensure the potential barrier is high enough to stop atoms from hitting the fiber. Our calculated trap depth is $\sim 100 \mu$K with the atoms sitting a distance of approximately 350 nm from the nanofiber.

Atoms are cooled into the two-color trap by first preparing a cloud of Cesium atoms using a Magneto-Optical Trap (MOT), which is overlapped with the nanofiber. The atoms undergo two stages of cooling to load them into the two-color trap. The first is a compression stage where the cooling laser is detuned by $\sim$~12 MHz over 120 ms, causing the atom cloud to compress due to the reduced scattering rate. The second is a polarization gradient cooling (PGC) stage where the cooling laser is detuned an additional 22 MHz over 25 ms and the anti-Helmholtz coils of the MOT are switched off. After this second stage, the atoms are measured to be cooled to $\sim 10~\mu$K, much less than the calculated trap depth of $\sim 100 \mu$K.

Once the trap is loaded, the cooling lasers are switched off and the transmission through the nanofiber is probed using a near-resonance laser pulse of 100 $\mu$s. The laser frequency is controlled using an AOM in double pass configuration and is scanned 40 MHz across the atom transition frequency. The output of the fiber is measured using a Single Photon Counting Module (SPCM).

\section*{Results}

\begin{figure}
\centering
	\includegraphics[width=0.9 \linewidth]{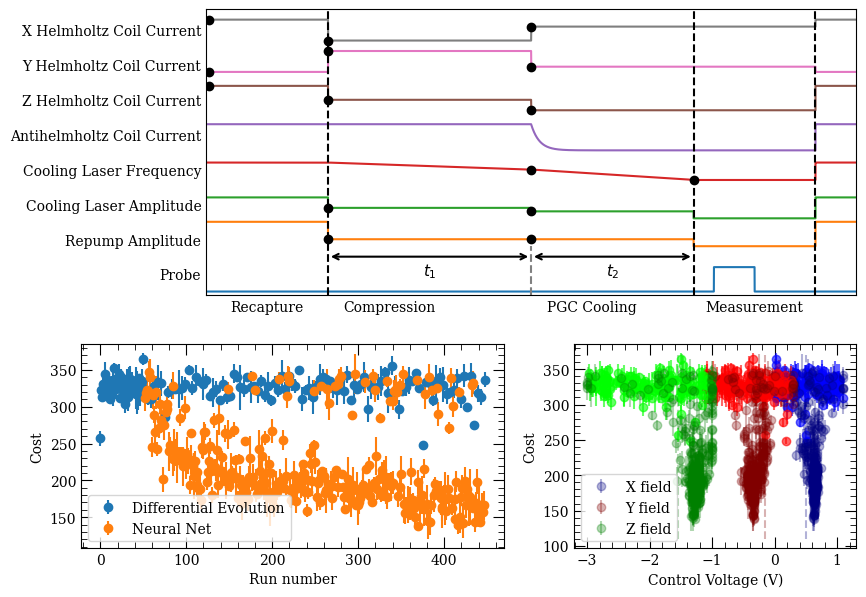} 
	\caption{{\bf(Top)} Cooling sequence used in the experiment. The dots show values which where adjusted by the machine learning algorithm, as were the compression ramp time $t_1$ and the PGC cooling time $t_2$. {\bf(Bottom Left)} Cost vs run number for an example machine learning session. The cost values are distinguished by whether the test parameters used were produced by the differential evolution algorithm or the neural net model. {\bf(Bottom Right)} Cost vs Control voltage for three parameters varied during a machine learning session. These were the X, Y and Z Helmholtz magnetic fields at the beginning of the compression cooling step. Darker colors indicate values that were produced by the neural net model and brighter colors the differential evolution algorithm. The dashed lines show the starting values of the parameters}
	\label{fig:MachineLearning}
\end{figure}

Because there were many parameters to be adjusted in the cooling sequence, we employed machine learning to help with the task of optimizing the number of atoms in the trap. This has previously been done by Gupta {\em et al.} \cite{gupta2022_machine_learner_opt} where they trapped Rb atoms. The Python package M-LOOP \cite{mloop,Wigley2016FastMLOO} was incorporated into our experimental control program for this purpose. The top of figure \ref{fig:MachineLearning} shows a graphical representation of the cooling sequence and all the parameters that were optimized by the machine learning algorithm. The transmission was used as the cost for the algorithm to minimize. 

The algorithm first obtained 50 measurements as training data, where each set of test parameters was determined using the differential evolution algorithm. After the 50 training runs, a neural net model was used to determine the next set of test parameters. A gaussian evolution algorithm might have been best to use due to the noisy nature of our data. However, because it scales with the cube of the number of measurements, it slows down quickly compared with the neural net model which scales linearly.

We found that the neural net model consistently converged toward locally optimal parameter sets. The largest number of parameters optimized at one time was 17. We were limited in how many measurements we could achieve in a session due to the limited time of our repump lock and generally each session was around 200-500 measurements. Optimization was initialized from configurations exhibiting measurable absorption. The bottom left of figure \ref{fig:MachineLearning} shows an example run, where the cost starts to reduce after the 50 training runs due to the neural net taking over. The neural net occasionally explores poorly predicted regions, which leads to deviations from this trend. However, the optimization algorithm continues toward the most optimal set of parameters.

The bottom right of figure \ref{fig:MachineLearning} shows the cost versus the test parameter value for 3 chosen parameters. These 3 parameters determine the position of the atom cloud, which tends to be spherically symmetric. This is why there seems to be only 1 minimum for the parameters, which is aligning the center of the cloud with the two-color trap.

The machine learner was run several times, using the set of test parameters with the lowest cost from the previous run. For the first run, the transmission measurement used the resonant frequency and measured directly after turning off all cooling lasers. At some point there will be enough atoms in the two-color trap such that the transmission will be lower than 1\%. This results in the machine learner being unable to optimize the cost any further as any increase in the number of atoms hardly changes the transmission. This is why for subsequent runs the time of the measurement was changed to happen at later times after turning off the cooling lasers.

The optimization procedure improved configurations with initial optical depths below 0.5. In one set of sessions, the starting transmission was 70\% on resonance, immediately after turning off the cooling lasers. The first session improved that transmission to 10\%. The measurement was then set to happen at 50 ms after turning off the cooling lasers, where the starting transmission was 45\% and improved to 20 \%. For the final session the measurement started at 95 ms after switching off cooling lasers, and the start and end transmissions were 60\% and 45\%. Each of these sessions had around 300 measurement points and the total time for this whole optimization was 4 hours. The final measured optical density measured here was 4.5, which is most likely limited due to the initial small MOT size as compared to our best measurements, but could also have been due to the algorithm optimizing to a minimum which had a lower optical density.

The frequencies and amplitudes of the trapping lasers did not change significantly over an optimization, it was mainly the magnetic fields values that were key. Most important was the interplay between setting the position of the atoms before turning off the anti-Helmholtz coil and the compensation fields for switching it off. There was likely a dependence between these two sets of fields, which was difficult to optimize by hand. In some cases, machine learning was applied just to these fields as these parameters were found to be the most important to increase the optical density.

\begin{figure}
\centering
	\includegraphics[width=0.9 \linewidth]{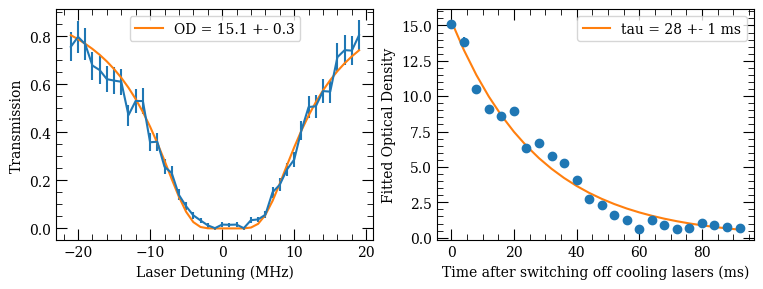} 
	\caption{{\bf(Left)} Transmission measurement through the fiber with atoms in the two-color trap. The fit is using equation \ref{eq:ODfit} for the optical density. {\bf(Right)} Lifetime measurement of the atoms in the trap. Derived from optical density measurements for a range of times after turning off the cooling lasers.}
	\label{fig:OpticalDensityAndDecay}
\end{figure}

After optimization of the cooling sequence parameters, the transmission was measured over a detuning of 40 MHz as shown in figure \ref{fig:OpticalDensityAndDecay}(left). The transmission minimum saturates due to the high optical density and hence number of atoms in the two-color trap. We can fit the transmission using an expression derived from the Beer-Lambert law where two-level systems are involved \cite{Raskop2020QuantumOptics,Grover2015Nanofibers}:
\begin{equation} \label{eq:ODfit}
    T(\Delta) = \exp{\left( - \frac{OD}{1 + 4((\Delta - \Delta_0)/\Gamma)^2}\right)}
\end{equation} 
where $\Delta$ is the detuning and $\Delta_0$ the resonant frequency, $\Gamma = 2\pi \times 5.22$ MHz is the Cs natural linewidth and $OD$ is the total optical depth. Our trap shows an optical depth of 15.1 $\pm$ 0.3.

Because we know our atoms sit a distance 350 nm away from the fiber, we can calculate the optical density per atom. This is given by the ratio of the resonant cross section of the atom to the effective area of the mode at the atom position. The single atom optical density will be \cite{Raskop2020QuantumOptics}
\begin{equation}
    OD_{1atom} = \frac{3 \lambda^3}{2 \pi} \frac{I(\bf{r}_{atom})}{P_{mode}}
\end{equation}
where $\lambda$ is the wavelength of the transition and $I(\bf{r}_{atom})$ is the intensity of the field at the atom position and can be calculated using the equations for the EM fields around an optical nanofiber for example from \cite{Nayak2018NanofiberQuantumPhotonics,Vetsch2011NanofiberInterface}. The power in the mode $P_{mode}$ ends up canceling out as it also appears in the calculation of the intensity. 

For our experiment we used a probe that had the same linear polarization as the trapping lasers. The result is that in our experiment the calculated $OD_{1atom} = 0.011$, and therefore we estimate our trap contains around 1400 atoms. This is smaller than in the 2010 work \cite{vetsch2010_optical_interface} which estimated around 2000 atoms in their trap, however, they had a trapping potential 4 times larger than ours. Because the atom wavefunction is likely to spread out in the azimuthal direction, we would expect the actual $OD_{1atom}$ to be related to an integral around the azimuthal direction and therefore we probably over estimate our $OD_{1atom}$.

Based on the distance the atom sits from the nanofiber, the theoretical single atom cooperativity is ~0.02 \cite{le_kien2005_spontaneous}. With 1400 atoms in the trap the collective cooperativity is ~28.

To measure the lifetime of atoms out of the trap, we measure the optical density at moments in time after turning off the cooling lasers. Fig.~\ref{fig:OpticalDensityAndDecay}(right) shows the decay of the optical density over time with a fitted exponential function. The decay constant has a value of 28 $\pm$ 1 ms. This does not compare well with the lifetime of 50 ms from Vetsch {\em et al.} \cite{vetsch2010_optical_interface}, which could be due to our polarization configuration. It does, however, compare favorably to the state insensitive trap result of 12 ms \cite{goban2012_state_insensitive_nanofiber}. 

To understand the observed lifetime of $28 \pm 1\,\mathrm{ms}$, we compare the relevant heating mechanisms with the trap energy scale ($\sim 100\,\mu\mathrm{K}$). The calculated recoil heating rate from off-resonant scattering of the trapping fields is approximately $1\,\mu\mathrm{K/s}$, too small to account for atom loss on this timescale. Furthermore, the vacuum-limited lifetime exceeds 1~s, excluding background gas collisions as the dominant mechanism. These considerations rule out fundamental limits and instead indicate technical heating.

In a harmonic trap, intensity or phase noise at frequencies near $2\omega_i$ may drive parametric excitation. With radial and axial trap frequencies of $\sim 600$ kHz and 1.02 MHz respectively, even small spectral components of laser noise near twice these frequencies can lead to exponential growth of the kinetic energy of the atoms. 
Previous nanofiber trap experiments have suggested intensity or phase fluctuations of the trapping fields as possible heating mechanisms.
In our configuration, the weak azimuthal confinement (as illustrated in Fig~\ref{fig:experimentalsetup}) 
further increases sensitivity to slow intensity gradients and technical drifts, allowing atoms to sample regions of reduced trap depth. 
The measured lifetime is therefore consistent with technical noise amplified by the anisotropic confinement of the trap.

\begin{figure}
\centering
	\includegraphics[width=0.9 \linewidth]{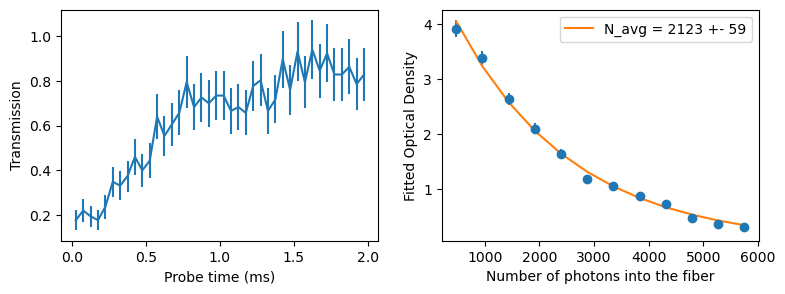} 
	\caption{{\bf(Left)} Transmission measurement on resonance through the fiber. {\bf(Right)} Fitted optical density vs number of photons into the trap. The fit is exponential and decay constant is an average number of photons 2120 $\pm$ 60.}
	\label{fig:Heating}
\end{figure}

Probe-induced heating provides a controlled comparison. When the probe laser is on, the trapped atoms are heated by the recoil of the absorbed photons. As a consequence, as shown in Fig.~\ref{fig:Heating} (left), the absorption decreases with the time the probe laser is on. In Fig.~\ref{fig:Heating}(right) we show that the optical depth decreases exponentially with the number of probe laser photons going into the fiber, with a characteristic scale of 2120 $\pm$ 60 photons. Assuming that each scattered photon contributes one recoil energy, the corresponding increase in energy is $\sim 210\,\mu\mathrm{K}$, comparable to the difference between the atomic temperature and trap depth. This agreement supports recoil-driven escape as the mechanism for probe-induced loss.

These results indicate that the present confinement time is limited by technical noise rather than intrinsic recoil or vacuum processes. Improving laser phase stability and enhancing azimuthal confinement would increase the usable interaction time while preserving the measured collective cooperativity ($C_N \approx 28$), thereby enabling exploration of collective emission and waveguide-QED dynamics over longer timescales.

\begin{table}[h]
    \centering
    \begin{tabular}{l|l|l|l|l|l}
    Work & Species & OD & \# Atoms & Lifetime & Trap type \\ \hline
         Vetch 2010 \cite{vetsch2010_optical_interface} & Cs & $\sim 13$ & $\sim 2000$ & 50 ms & Stronger trap \\
    Goban 2012 \cite{goban2012_state_insensitive_nanofiber} & Cs & 66 & $\sim 600$ & 12--140 ms & state-insensitive \\
    Gupta 2022 \cite{gupta2022_machine_learner_opt} & Rb & 5.3 & 450 & ? & near-resonant\\
    This work & Cs & 15.1 & $\sim 1400$ & 28 ms & shallow trap 
    \end{tabular}
    \caption{Comparison with recent literature}
    \label{tab:one}
\end{table}
In table~\ref{tab:one} we compare our results with the recent literature. 
Despite operating in a relatively shallow and weakly confined trapping regime, the achieved optical depth and atom number are comparable to earlier nanofiber trap implementations.





\section*{Conclusion}

In conclusion, we have successfully implemented a two-color dipole trap, resulting in the trapping of an estimated 1400 atoms around our nanofiber with a decay constant of 28 $\pm$ 1 ms. Machine learning was an invaluable tool for optimizing the number of atoms in the trap, with the python package MLOOP \cite{mloop} providing an easy way to integrate this into our experimental control software. The demonstrated combination of high optical depth and automated optimization establishes this platform as a useful testbed for collective atom–waveguide interactions and cavity-enhanced nanofiber QED.

\bibliography{sample}

\section*{Acknowledgements}

This work was supported by the Marsden Fund under grant number UOA2127. The authors would like to thank Dr Dylan J Brown for his useful guidance on getting our experiment working.

\section*{Data Availability}
The data that support the findings of this study are available from the corresponding author upon reasonable request. 

\section*{Author Declarations}
\subsection*{Conflict of interest}
The authors have no conflicts of interest to disclose.
\section*{Author contributions}
M.H. conceived the experiment(s),  W.C. and M.S. conducted the experiment(s), W.C. analysed the results.  All authors reviewed the manuscript.

\end{document}